\documentclass[12pt]{iopart}			
\usepackage{graphicx}

\begin{document}

\title[Elliptic flow at high transverse momentum in ALICE]{Elliptic flow at high transverse momentum in Pb-Pb collisions at $\sqrt{s_{NN}} = 2.76$~TeV with the ALICE experiment}

\author{A.~Dobrin for the ALICE Collaboration}
\address{Department of Physics and Astronomy, Wayne State University, 666 W. Hancock, Detroit, Michigan 48201}
\ead{alexandru.florin.dobrin@cern.ch}

\begin{abstract}
An observable that can be used to better constrain the mechanism responsible for the parton energy loss is the elliptic azimuthal event anisotropy, $v_2$. We report on measurements of $v_2$ for inclusive and identified charged particles in Pb-Pb collisions at $\sqrt{s_{NN}} = 2.76$~TeV recorded by the ALICE experiment at the LHC. $v_2$ is presented for a wide range of particle transverse momentum up to $p_T=20$~GeV/c. The particle $v_2$ is finite, positive and approximately constant for $p_T>8$~GeV/c. The proton $v_2$ is higher than that of the pion up to about $p_T=8$~GeV/c. The results are compared to the measurements at lower energy reported by RHIC experiments.
\end{abstract}

\section{Introduction}

The main goal of heavy-ion physics is to study the new phase of matter predicted by Quantum Chromodynamics to exist at high energy density, the so-called Quark-Gluon Plasma (QGP). An important observable used to characterize its properties is the anisotropic flow~\cite{Ollitrault:1992bk}. The elliptic flow, $v_2$, is the second harmonic coefficient in a Fourier series expansion of the particle azimuthal distribution measured with respect to the reaction plane~\cite{Voloshin:1994mz}. The reaction plane is spanned by the beam axis $z$ and the impact parameter vector. There are two main issues not related to the initial geometry that complicate the flow measurements: non-flow (other sources of azimuthal correlations) and event-by-event flow fluctuations.

The large elliptic flow observed at RHIC showed that the matter created in heavy-ion collisions behaves as a nearly ``perfect'' fluid whose constituent particles interact very strongly~\cite{Adams:2005dq}. The ALICE Collaboration reported that, compared to RHIC energies the differential elliptic flow  ($0.2<p_T<5$~GeV/c) does not change within uncertainties at LHC energies, but the integrated elliptic flow increases by about 30\%~\cite{Aamodt:2010pa}. 

In this paper we extend the $p_T$ reach up to 20~GeV/c for unidentified charged particles $v_2$ in ALICE; also charged pions and (anti)protons $v_2$ were measured up to $p_T=20$~GeV/c.

\section{Data taking and results}

The data recorded by the ALICE experiment~\cite{Aamodt:2008zz} during the November-December 2010 LHC heavy-ion run were used in this paper. The Time Projection Chamber (TPC) was used for tracking, particle identification and reconstruction of the event plane. Other detectors, e.g. Zero Degree Calorimeter (ZDC) and VZERO, were employed to determine the event plane. The event sample was collected with a dedicated minimum bias trigger. In order to remove beam induced background events, an offline event selection is applied. Furthermore, only events with a reconstructed primary vertex within $\pm 10$ cm along the beam axis are selected for this analysis. Such a selection leads to around 14.5 million events. Only charged particles with $p_T>0.2$~GeV/c are selected in the pseudo-rapidity range $|\eta|<0.8$. A Glauber model fit to the distribution of the summed amplitudes in the VZERO scintillator detectors was used to estimate the centrality of the collision~\cite{Toia:QM11}.

\begin{figure}[tp]
  \centering
  \includegraphics[keepaspectratio, width=0.49\columnwidth]{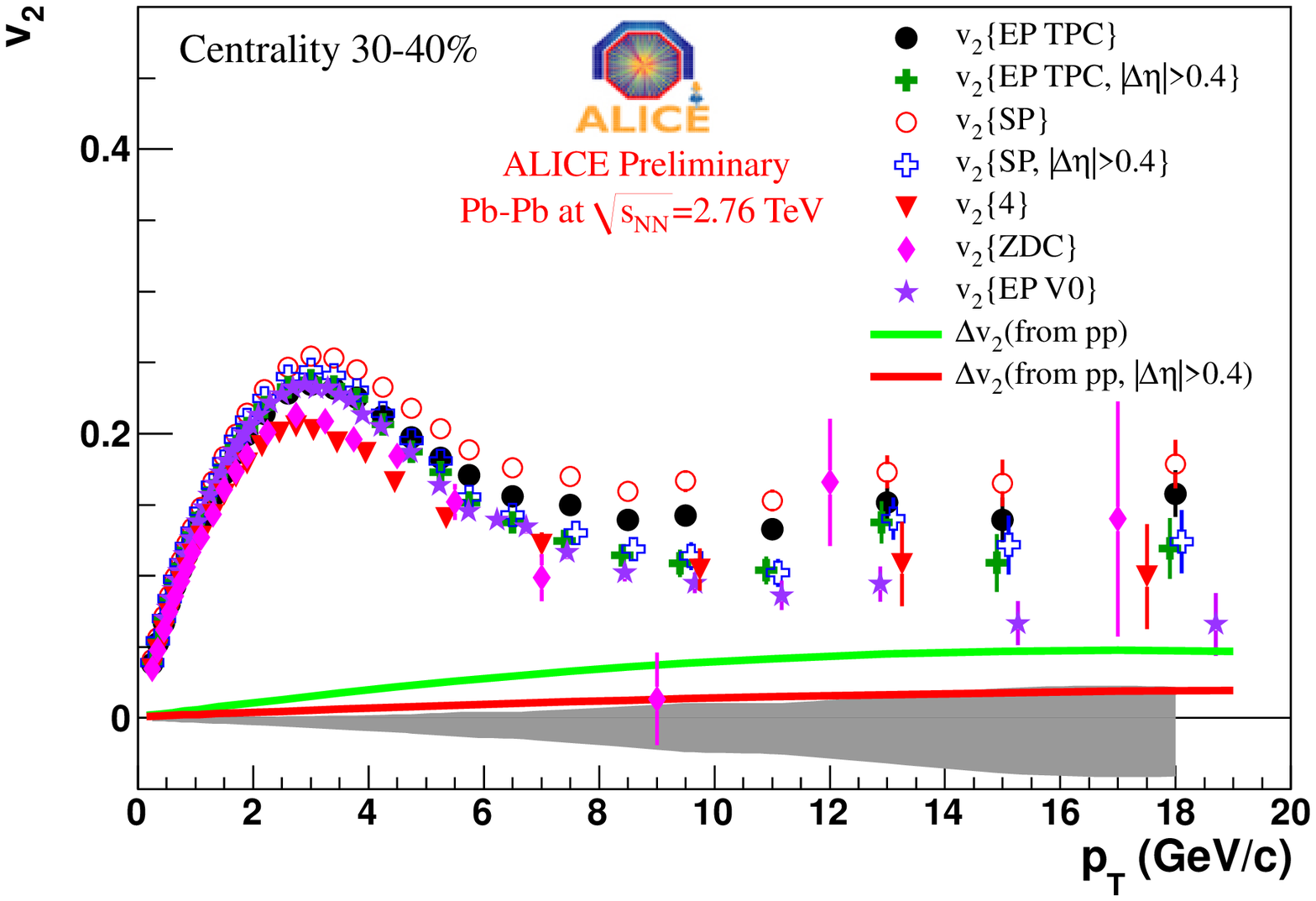}
  \includegraphics[keepaspectratio, width=0.49\columnwidth]{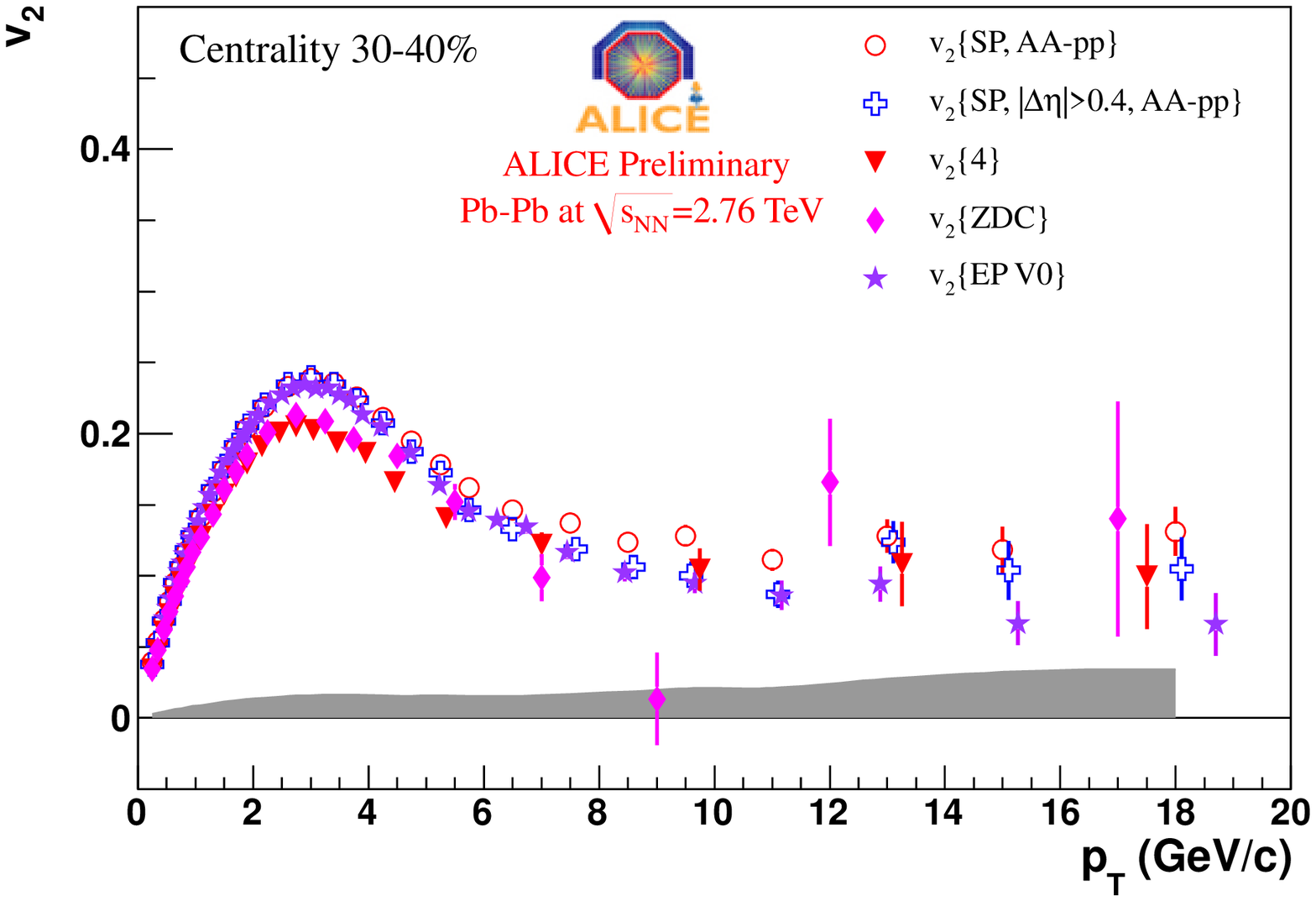}
  \caption{(color online) Left: $v_2(p_T)$ for Pb-Pb collisions in the centrality bin 30-40\% from various analysis techniques. The solid lines are the estimated non-flow correction factors from pp data using the $\langle u^*Q \rangle$ method without any $\eta$ gap (green) and with an $\eta$ gap, $|\Delta\eta|>0.4$ (red). The gray band represents the systematic uncertainty for the $|\Delta\eta|>0.4$ methods. Right: Comparison between $v_2\{\mathrm{SP}\}$ corrected for the remaining non-flow using pp estimates and $v_2\{4\}$, $v_2\{\mathrm{ZDC}\}$, $v_2\{\mathrm{EP~V0}\}$. The systematic uncertainty is depicted by the gray band.}
  \label{fig:v2_mth_comp}
\end{figure}

In this analysis several experimental methods were investigated: the event plane method, and two and four particle correlations. For the first, the event plane was determined from TPC tracks ($v_2\{\mathrm{EP~TPC}\}$) and VZERO signals ($v_2\{\mathrm{EP~V0}\}$). In the case of two particle correlations the scalar product method ($v_2\{\mathrm{SP}\}$) was employed, while for four particle correlations the analysis was performed using the cumulant method ($v_2\{4\}$). Furthermore, the charged particle elliptic flow is also calculated with the reaction plane determined from directed flow of neutral spectators measured with the ZDC ($v_2\{\mathrm{ZDC}\}$). For a detailed description of the methods and notation see~\cite{Voloshin:2008dg}.

An important part of the analysis is devoted to the removal of non-flow contributions. This is done by introducing an $\eta$ gap ($|\Delta\eta|> 0.4$) and correlating particles from $\eta \in (-0.8, -0.2)$ with particles from $\eta \in (0.2, 0.8)$ and vice-versa. When the $\eta$ gap is increased (using VZERO to determine the event plane), the non-flow is suppressed even more. We estimate the remaining non-flow contribution by applying the $\langle u^*Q \rangle$ method on pp data at the same energy, in which the correction is given by $\frac{\langle u^*Q \rangle^{pp}} {M \langle v_2 \rangle}$, where $\langle v_2 \rangle$ is an average flow for each centrality bin and $M$ is the mean multiplicity~\cite{Voloshin:2008dg}.

Figure~\ref{fig:v2_mth_comp} (left) shows the $p_T$-differential elliptic flow measured for centrality bin 30-40\% using various analysis techniques. The estimated non-flow corrections are depicted by the solid lines. The results differ as expected since the methods have different sensitivity to flow fluctuations and non-flow. The right panel of Fig.~\ref{fig:v2_mth_comp} presents comparisons between $v_2\{\mathrm{SP}\}$ corrected for non-flow (denoted by $v_2\{\mathrm{SP, AA-pp}\}$) and $v_2\{4\}$, $v_2\{\mathrm{ZDC}\}$, $v_2\{\mathrm{EP~V0}\}$. A good agreement between $v_2\{\mathrm{SP}\}$ and $v_2\{\mathrm{EP~V0}\}$ is found over the $p_T$ range considered; the $v_2\{4\}$ and $v_2\{\mathrm{ZDC}\}$ results deviate from the $\langle v_2 \rangle$ for $p_T<7$~GeV/c due to flow fluctuations. 

\begin{figure}[tp]
  \begin{center}  
    \includegraphics[keepaspectratio, width=0.55\columnwidth]{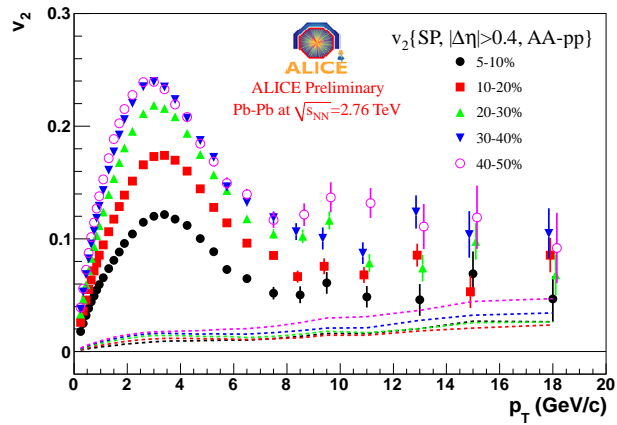}
    \caption{(color online) $v_2(p_T)$ for Pb-Pb collisions for various centralities from the $v_2\{\mathrm{SP},|\Delta\eta|>0.4, \mathrm{AA}-\mathrm{pp}\}$ method. The data points for $p_T>10$~GeV/c are shifted slightly in $p_T$ for visibility. The estimated non-flow corrections from pp are included in the systematic errors (represented by dashed lines).}
    \label{fig:v2_cent}
  \end{center}
\end{figure}

Figure~\ref{fig:v2_cent} presents $v_2(p_T)$ for different centrality classes from $v_2\{\mathrm{SP}, |\Delta\eta|>0.4, \mathrm{AA}-\mathrm{pp}\}$ method. $v_2$ at high transverse momentum ($p_T>8$~GeV/c) is finite and positive; it reaches constant values which increase from central to mid-peripheral events.

Identified particle (PID) flow is a useful tool for understanding the different $v_2$ regimes. PID $v_2$ gives insight into the hydrodynamical behavior of the created matter and can help determine the region where the coalescence picture appears to be important. 

At intermediate/high $p_T$ ($3<p_T<20$~GeV/c) PID of charged hadrons can be done by employing the ionization energy loss, $dE/dx$, in the TPC. The TPC PID is based on the observable $\Delta_{\pi} = dE/dx - \langle dE/dx \rangle_{\pi}$, where $\langle dE/dx \rangle_{\pi}$ is the expected energy loss for a pion. Particles with $1<\Delta_{\pi}<7$ and $-25<\Delta_{\pi}<-14$ were identified as pions and (anti)protons, respectively. The estimated contamination was found to be below 1\% for pions and below 15\% for protons, respectively in the measured $p_T$ range.

\begin{figure}[tp]
  \centering
  \includegraphics[keepaspectratio, width=0.49\columnwidth]{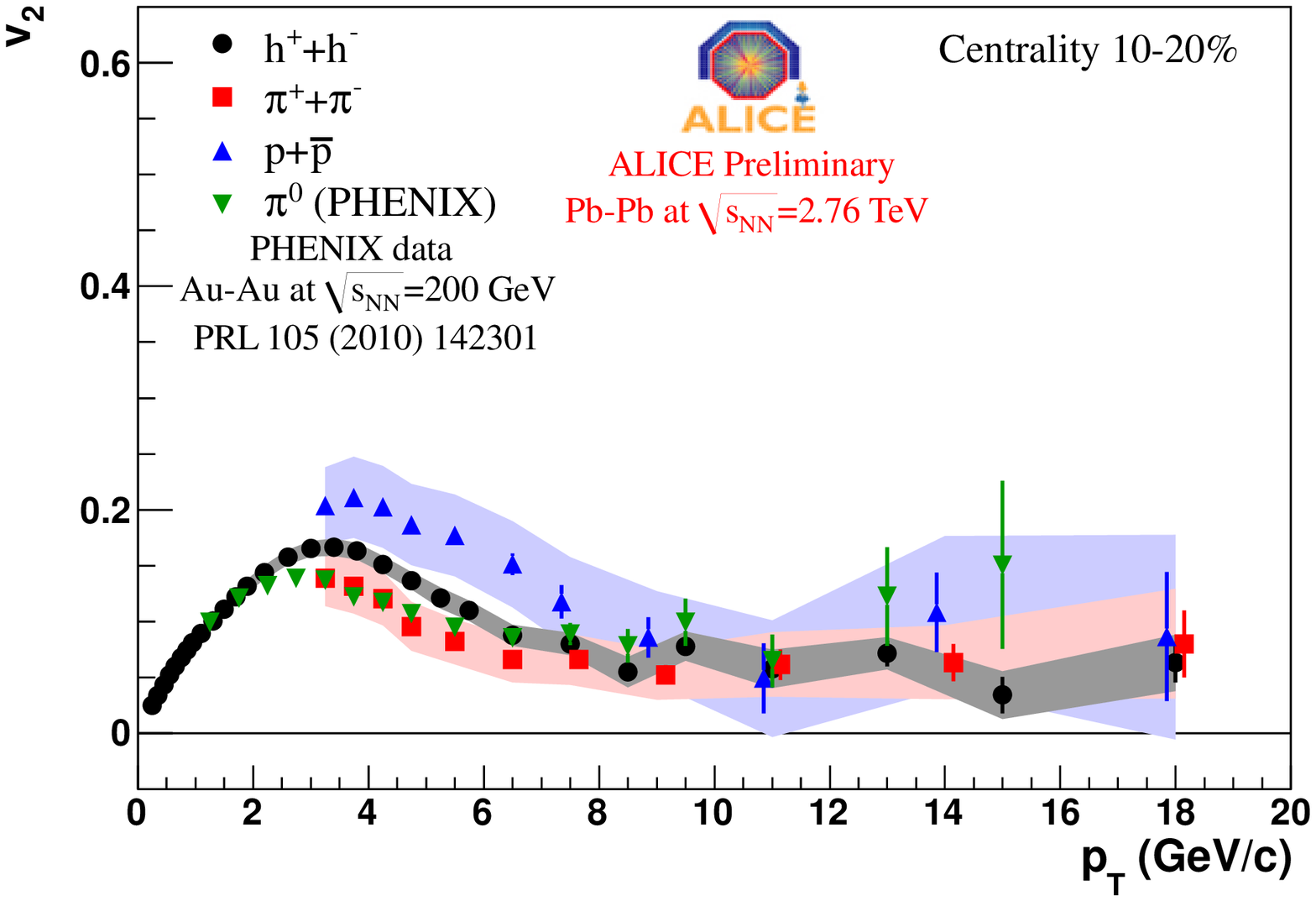}
  \includegraphics[keepaspectratio, width=0.49\columnwidth]{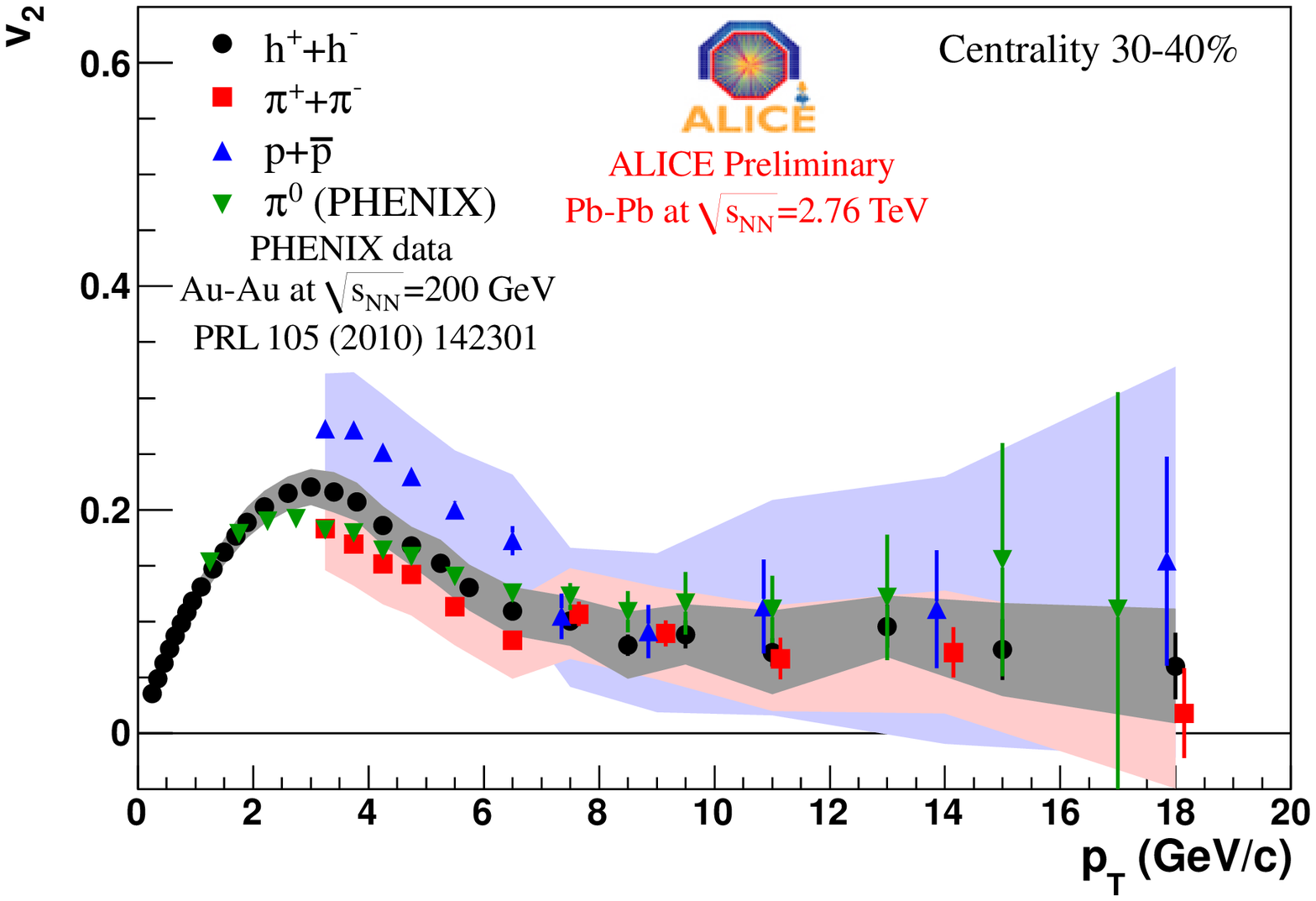}
  \caption{(color online) Pion and proton $v_2(p_T)$ compared with unidentified charged particles $v_2(p_T)$ for centrality bins 10-20\% (left) and 30-40\% (right) from $v_2\{\mathrm{EP~TPC},|\Delta\eta|>0.4\}$ method. The results are corrected for the remaining non-flow using data from the centrality bin 70-80\%. Data points are shifted for visibility. PHENIX $\pi^0$ measurements for Au-Au at $\sqrt{s_{NN}}=200$~GeV are also shown.}
  \label{fig:v2_pid}
\end{figure}

Figure~\ref{fig:v2_pid} shows the charged pion and proton $v_2$ for centrality bins 10-20\% (left) and 30-40\% (right) from $v_2\{\mathrm{EP~TPC},|\Delta\eta|>0.4\}$ method. The remaining non-flow was corrected for by using data from centrality bin 70-80\% scaled by multiplicity:
\begin{equation}
v_{2,cent}^{corr} = v_{2,cent} - v_{2,70-80\%}\frac{M_{70-80\%}}{M_{cent}}.
\end{equation}
The proton $v_2$ is higher than that of the pion at intermediate $p_T$, while the results start to overlap within systematic uncertainties for $p_T>8$~GeV/c. An agreement with $\pi^0$ $v_2$ measured by PHENIX in Au-Au collisions at $\sqrt{s_{NN}}=200$~GeV is found within uncertainties.

\section{Summary}

Charged particle elliptic flow (both for inclusive and identified charged particles) has been measured up to $p_T=20$~GeV/c. We found that unidentified charged particle $v_2$ is finite, positive and approximately constant at $p_T>8$~GeV/c. Proton $v_2$ is higher than that of the pion up to about $p_T= 8$~GeV/c.

\end{document}